\documentclass[a4paper,fleqn,useAMS,usenatbib]{mnras}

\usepackage{amssymb}
\usepackage{amsmath}
\usepackage{times}
\usepackage{graphicx}
\usepackage{pgf}

\def\msh{PSR\,B1509--58}

\def\axj{AX J1838.0--0655}
\def\igr{IGR J1849.0--0000}

\def\psra{PSR\,J1846--0258} 
\def\psrb{PSR\,J1119--6127} 
\def\psrc{PSR\,J1124--5916} 
\def\psrn{PSR\,J1208--6238}

\newcommand{\gr}{$\gamma$}
\newcommand{\xmm}{{\it XMM-Newton}}
\newcommand{\fermi}{{\it Fermi}}
\newcommand{\integral}{{\it INTEGRAL}}
\newcommand{\cxo}{{\it Chandra}}
\newcommand{\rxte}{{\it RXTE}}

\newcommand{\cgro}{{\it CGRO}}
\newcommand{\nustar}{{\it NuSTAR}}

\newcommand{\swift}{{\it Swift}}

\title[\fermi\ LAT detection of \psra]{The \fermi-LAT detection of magnetar-like pulsar \psra\ at high-energy gamma-rays}
\author[L. Kuiper, W. Hermsen and A. Dekker]{L. Kuiper$^{1}$\thanks{E-mail: L.M.Kuiper@sron.nl (LK)}, W. Hermsen$^{1,2}$ and A. Dekker$^{1}$\\
$^{1}$SRON-Netherlands Institute for Space Research, Sorbonnelaan 2, 
3584 CA, Utrecht, The Netherlands\\
$^{2}$Astronomical Institute ``Anton Pannekoek", University of Amsterdam, Science Park 904, 1098 XH Amsterdam, The Netherlands\\
}

\begin{document}

\date{Accepted 2017 November 30. Received 2017 November 30 ; in original form 2017 August 29}

\pagerange{\pageref{firstpage}--\pageref{lastpage}} \pubyear{2016}

\maketitle

\label{firstpage}

\begin{abstract}
{We report the detection of the pulsed signal of the radio-quiet magnetar-like pulsar \psra\ in the high-energy \gr-ray data of the \fermi\ Large Area Telescope (\fermi\ LAT).
We produced phase-coherent timing models exploiting \rxte\ PCA and \swift\ XRT monitoring data for the post- (magnetar-like) outburst period from 2007 August 28 to 2016 September 4,
with independent verification using \integral\ ISGRI and \fermi\ GBM data. Phase-folding barycentric arrival times of selected \fermi\ LAT events from \psra, resulted in a $4.2\sigma$ 
detection (30--100 MeV) of a broad pulse consistent in shape and aligned in phase with the profiles that we measured with \swift\ XRT (2.5--10 keV), \integral\ ISGRI (20--150 keV) 
and \fermi\ GBM (20--300 keV). The pulsed flux (30--100 MeV) is $(3.91\pm0.97)\times 10^{-9}$ photons cm$^{-2}$ s$^{-1}$ MeV$^{-1}$. 
Declining significances of the \integral\ ISGRI 20--150 keV pulse profiles suggest fading of the pulsed hard X-ray emission during the post-outburst epochs.
We revisited with greatly improved statistics the timing and spectral characteristics of \msh\ as measured with the \fermi\ LAT. 
The broad-band pulsed emission spectra (from 2 keV up to GeV energies) of \psra\ and \msh\ can be accurately described with similarly curved shapes, with maximum luminosities 
at $3.5 \pm 1.1$ MeV (\psra) and $2.23 \pm 0.11$ MeV (\msh). We discuss possible explanations for observational differences between 
\fermi\ LAT detected pulsars that reach maximum luminosities at GeV energies, like the second magnetar-like pulsar \psrb, and pulsars with maximum luminosities at MeV energies, 
which might be due to geometric differences rather than exotic physics in high-B fields.
}
\end{abstract}



\section{Introduction}

\psra\  was discovered as a 0.3 s X-ray pulsar \citep{gotthelf2000}, 
located at the centre of supernova remnant (SNR) Kes 75 \citep{helfand2003}.
It is a young (characteristic age $\tau \sim 723$ yr) radio-quiet \citep[e.g.][]{archibald2008} rotation-powered pulsar, with $P \sim 324$ ms.
Its surface magnetic field strength of $4.9 \times 10^{13}$ G is above the quantum critical field strength of 
$4.413 \times 10^{13}$ G. 
\rxte\ monitoring before 2006 June showed that \psra\/ behaved as a very stable rotator \citep{livingstone2006}.
With \integral\ a hard X-ray point-source (pulsar plus the surrounding diffuse pulsar wind nebula) and pulsed emission were detected.
For details about the pulsed and total high-energy spectrum across the $\sim$ 2-300 keV band we refer to \citet{kuiper2009}.

Most intriguingly,  \psra\ was the first rotation-powered pulsar that exhibited magnetar-like behaviour, starting with  a dramatic brightening of the pulsar in
\cxo\ observations of Kes 75 during 2006 June 7-12 \citep{kumar2008}.  Furthermore, 
\citet{gavriil2008} reported that this radiative event lasted for about 55 d
and discovered five short magnetar-like bursts during the outburst. In a followup study, \citet{kuiper2009} discovered that the onset of the radiative event was accompanied by 
a strong glitch in the rotation behaviour of the pulsar. 
Using multi-year \rxte\ and \integral\ observations, they concluded that \psra\ exhibited before its magnetar-like outburst
over many years a very stable behaviour, both temporally and spectrally, and
continued after its outburst again as a young stable energetic rotation-powered X-ray pulsar. 
At higher energies, \citet{parent2011} did not detect the pulsed signal of \psra\ for energies 
above 100 MeV, analysing about 20 months of \fermi\ LAT data.

Recently, a magnetar-like outburst has been detected from a second high-B rotation-powered pulsar, namely the radio pulsar \psrb\ \citep[][$\tau \sim 1.6$ ky]{camilo2000},
with very similar characteristics as observed during the outburst of \psra. Particularly, an equally strong spin-up glitch followed by a radiative outburst accompanied 
by a few short magnetar-like bursts \citep{archibald2016}. 
Different from \psra, \citet{parent2011} reported \psrb\ to be a \fermi\ high-energy gamma-ray source with maximum luminosity 
at GeV energies, typical for the population of \fermi -detected gamma-ray pulsars. 
This is in contrast to the spectrum of the young high-B-field soft gamma-ray pulsar, \msh\ (aka PSR J1513-5908; $\tau \sim 1.6$ ky), 
which has after the Crab pulsar the highest flux at hard X-ray energies, and has been detected up to the GeV band of the \fermi\ LAT, but reaches its maximum 
luminosity at MeV energies \citep{kuiper2015}.

\psra\ and \msh\ have very similar spectral shapes of the pulsed emissions in the X-ray band above 2 keV (also the pulse shapes are similar). If the spectral shapes remain similar across the full X-ray / gamma-ray band,
then a 5--10 $\times$ smaller 30-100 MeV flux is expected for \psra\ relative to that of \msh. However, if the gamma-ray luminosity peaks at GeV energies, like detected
for the second "magnetar-like" pulsar \psrb\, then a stronger high-energy gamma-ray source should be observed. 

Currently, there are more than 8 years of \fermi\ LAT data available and the events are reconstructed adopting a new strategy 
called `Pass 8', which greatly enhances the sensitivity at lower gamma-ray energies ($\la 300$ MeV), compared to e.g. earlier attempts to detect a pulsed signal from 
\psra\ \citep{parent2011}. 
The combination of long exposure times and the enhanced sensitivity at lower gamma-ray energies allow deeper quests to the 
expected high-energy gamma-ray spectral tail of the pulsed emission of \psra. However, no radio-ephemerides can be derived for this pulsar, what would
have made pulse-phase folding of the arrival times of the \fermi\ events a routine task. But, \psra\ is a relatively bright X-ray pulsar, and, because of its unique properties, it is 
almost continuously monitored at X-rays since its discovery as a pulsar in 1999. First with the PCA aboard \rxte\ till 
2012 January and beyond with the XRT instrument aboard \swift. These X-ray monitoring observations allow the construction of 
reliable timing models.

In this work we aim to detect pulsed high-energy gamma-ray emission from \psra\ in the \fermi\ LAT passband using all available (Pass-8) data. 
For this purpose we had to generate phase-coherent timing models (ephemerides) for \psra\ using \rxte\ PCA and \swift\ XRT monitoring observations
across the post-outburst/giant-glitch period MJD 54340-57635 (2007 August 28 -- 2016 September 4). 
Because of the complexity of the ephemeris construction from \swift\ XRT data (period beyond 2012 January) we verified the 
resulting timing models using (independent) \integral\ ISGRI and \fermi\ GBM data. 
After the validity checks we applied the appropriate
timing models in \fermi\ LAT data-folding procedures. Furthermore, we derived the total pulsed hard-X-ray to gamma-ray spectrum of
\psra\ for comparison with those of \psrb\ and \msh. For the latter comparison, we revisited and significantly improved the published high-energy gamma-ray spectrum of \msh\, 
by exploiting the higher sensitivity of \fermi\ LAT Pass-8 data and additional observations.

In Section 2 we present all X-ray and gamma-ray instruments used in this analysis. Section 3 describes the tedious analysis of the multi-instrument X-ray 
data to arrive at phase-coherent timing models that are used in Section 4 for the \fermi\ LAT timing analysis. In Section 4 we also (re)derive the pulsed hard-X-ray -- gamma-ray
spectra of \psra\ and \msh. Finally, in Sections 5 and 6 we summarize and discuss our results. 

\section{Instruments} 

In this section we briefly present the instruments aboard \rxte, \swift, \integral\, and \fermi, which 
we used in this work. For their general characteristics we refer to the descriptions by the respective instrument teams,
referenced below.

\subsection{\rxte\ PCA}
\label{pca_char}
\rxte\ was launched on 1995 December 30 and ended its observations on 2012 January 5.
We used its PCA \citep{jahoda96} which consists of five collimated Xenon proportional 
counter units (PCUs), sensitive to 
photons with energies in the range 2-60 keV. All PCA data used in this study have been collected from observations in {\tt GoodXenon} mode 
allowing high-time resolution ($0.9\umu$s) analyses in 256 spectral bins.
 
\subsection{\swift\ XRT}
\label{instr_swift}
The \swift\ satellite \citep{gehrels04} was launched on 2004 November 20.
 \swift\ carries three co-aligned instruments: the wide-field coded aperture mask Burst Alert Telescope (BAT; 15-150 keV), the narrow field 
($23\farcm 6 \times 23\farcm 6$) grazing incidence Wolter-1 X-Ray Telescope (XRT; 0.2-10 keV) and 
the Ultraviolet/Optical Telescope (UVOT). We used data of \psra\ gathered 
by the XRT \citep{burrows05} from regular monitoring observations that commenced on 2011 July 25.
The XRT has several operation modes of which we only used the Windowed-Timing (WT) mode with a time resolution of 1.7675 ms, 
amply sufficient for pulse timing studies of \psra. 

\subsection{\integral\ ISGRI}
\label{instr_integral}
The \integral\ spacecraft \citep{winkler03}, launched on 2002 October 17, carries two main 
$\gamma$-ray instruments: a high-angular-resolution imager IBIS \citep{ubertini03} and
a high-energy-resolution spectrometer SPI \citep{vedrenne03}. 
In this work, guided by sensitivity considerations, we only used data recorded by the \integral\ 
Soft Gamma-Ray Imager ISGRI \citep{lebrun03}, the upper detector system of IBIS, sensitive 
to photons with energies in the range $\sim$15 keV -- 1 MeV (effectively about 300 keV).
The timing accuracy of the ISGRI time stamps recorded on board is about $61\umu$s. The time 
alignment between \integral\ and \rxte\ is better than $\sim 50\umu$s, verified using data 
from simultaneous \rxte\ and \integral\ observations of the accretion-powered millisecond pulsar 
IGR J00291+5934 \citep{falanga05}.

\subsection{\fermi}
\label{instr_fermi}
The \fermi\ gamma-ray space telescope was launched on 2008 June 11. It comprises two science instruments, the Large Area Telescope (LAT), sensitive 
to gamma-rays with energies between $\sim 20$ MeV and $300$ GeV, and the Gamma-ray Burst Monitor (GBM) covering the $\sim 8$ keV -- $40$ MeV energy band.
The spacecraft orbits the Earth in about 96.5 min at a height of about 550 km above the Earth's surface. After a checkout phase nominal science operations
started on 2008 August 4 with a one year all sky survey. Till 2013 December the LAT scanned the sky, providing all-sky coverage every two orbits. Since then the observing 
strategy has been modified combining sky survey observations with pointed observations including target-of-opportunity observations.

\subsubsection{Fermi LAT}
\label{instr_lat}
The Large Area Telescope aboard \fermi\ is an imaging, wide field-of-view (FoV $\sim$2.4 sr), high-energy \gr-ray telescope, covering the
energy range from below 20 MeV to more than 300 GeV \citep{atwood2009}. It is a pair-conversion telescope with a precision tracker and calorimeter, 
each consisting of a $4 \times 4$ array of 16 modules, a segmented anticoincidence detector that covers the tracker array,
and a programmable trigger and data acquisition system. The time stamps of the registered events are accurately ($\la 1 \umu$s ) derived from GPS clocks aboard the 
satellite. Since 2015 June 24, LAT events are available from a new event reconstruction and event selection strategy called Pass 8, allowing better sensitivity and
acceptance at lower energies than previous reconstructions \citep{atwood2013a,atwood2013b}.

\subsubsection{Fermi GBM}
\label{instr_gbm}
The Gamma-ray Burst Monitor \citep{bissaldi2009,meegan2009} comprises a set of 12 sodium iodide (NaI(Tl)) detectors sensitive across the 8 keV 
to 1 MeV band, and a set of 2 bismuth germanate (BGO) detectors covering the 150 keV to 40 MeV band, and so overlapping with the \fermi\ LAT passband. 
The set of non-imaging detectors provides a continuous view on each unblocked (by Earth) hemisphere.  During the first four years of the
operations, the timing accuracy was insufficient to perform timing studies for fast ($P < 512$ ms) energetic 
pulsars. However, since 2012 November 26 (MJD 56257), the GBM is operated in a nominal data-taking mode that provides time-tagged 
events (TTE) with $2\umu$s precision, synchronized to GPS every 
second, in 128 spectral channels, allowing now detailed timing studies at milli-second accuracies. For this work we exploited this much improved timing capability.


\begin{table*}
\caption{Phase-coherent post-outburst ephemerides for \psra\ as derived from \rxte\ PCA and \swift\ XRT (monitoring) data covering the time period
MJD 54340--57635 (2007 August 28 -- 2016 September 4).}
\label{eph_table}
\begin{center}
\begin{tabular}{lccclllccc}
\hline
Entry$^{a}$ &  Start &  End  &   t$_0$, Epoch   & \multicolumn{1}{c}{$\nu$}   & \multicolumn{1}{c}{$\dot\nu$}      & \multicolumn{1}{c}{$\ddot\nu$}                  & RMS & $\Phi_{0}$  & Validity range\\
 \#   &  [MJD] & [MJD] &     [MJD,TDB]    & \multicolumn{1}{c}{[Hz]}    & \multicolumn{1}{c}{$\times 10^{-11}$ Hz s$^{-1}$}  & \multicolumn{1}{c}{$\times 10^{-21}$ Hz s$^{-2}$}  &     &             &  (days)   \\
\hline\hline
1$^{b}$& 54340 & 54440 & 54340.0     & 3.064\,967\,014(15)  & -6.699\,9(7)              &   27(2)                  & 0.032    & 0.7992 &  101\\
\multicolumn{10}{l}{\dotfill}\\
\vspace{-0.25cm}\\
2           & 54559 & 55070 & 54559.0     & 3.063\,702\,807(2)   & -6.671\,56(2)             &   3.130(9)               & 0.036    & 0.4319 &  512\\
3           & 55056 & 55175 & 55056.0     & 3.060\,840\,893(19)  & -6.665\,1(8)              &   12.3(1.6)              & 0.050    & 0.4575 &  120\\
4           & 55154 & 55348 & 55222.0     & 3.059\,885\,948(3)   & -6.656\,45(4)             &   1.83(29)               & 0.033    & 0.6181 &  195\\
5           & 55348 & 55541 & 55494.0     & 3.058\,322\,717\,9(6)& -6.648\,75(7)             &   2.5(7.8)               & 0.040    & 0.4528 &  194\\
6           & 55488 & 55906 & 55811.0     & 3.056\,502\,936(2)   & -6.639\,46(2)             &   3.58(2)                & 0.033    & 0.7669 &  419\\
\multicolumn{10}{l}{\dotfill}\\
\vspace{-0.25cm}\\
7$^c$ & 55588 & 56185 & 55811.0     & 3.056\,502\,936\,3(7)& -6.639\,410(6)            &   3.67(1)                & 0.034    & 0.7713 &  598\\
\multicolumn{10}{l}{\dotfill}\\
\vspace{-0.25cm}\\
8$^d$& 56338 & 56967 & 56652.0     & 3.051\,688\,259(5)   & -6.613\,45(3)             &   3.28(7)                & .....    & 0.5898 &  630\\
9           & 56940 & 57458 & 57199.0     & 3.048\,566\,201(6)   & -6.598\,66(3)             &   3.17(9)                & .....    & 0.6503 &  519\\
10          & 57267 & 57635 & 57451.0     & 3.047\,130\,268(11)  & -6.591\,06(9)             &   3.71(37)               & .....    & 0.6461 &  369\\
\hline
\multicolumn{10}{l}{{\it Notes.}$^{a}$ Entries 1--6 are \rxte\ PCA ToA based, entry 7 is based on a combination of \rxte\ PCA and \swift\ XRT ToA's, and entries 8--10 are \swift\ XRT}\\
\multicolumn{10}{l}{\,\,\, based using a {\tt Simplex} optimization algorithm. Solar system planetary ephemeris DE200 has been used in the barycentering process.}\\
\multicolumn{10}{l}{$^b$ This entry provides an update on entry 9 of Table 2 of \citet{kuiper2009}}\\
\multicolumn{10}{l}{$^c$ Combined \rxte\ PCA / \swift\ XRT ephemeris bridging the 2011/2012 data gap}\\
\multicolumn{10}{l}{$^d$ An equivalent ephemeris, based solely on \swift\ XRT data, is shown in Table 1 of \citet{archibald2015}}\\

\end{tabular}
\end{center}
\end{table*}


\begin{figure*}
  \begin{center}
     \includegraphics[width=8cm,height=14cm,angle=90]{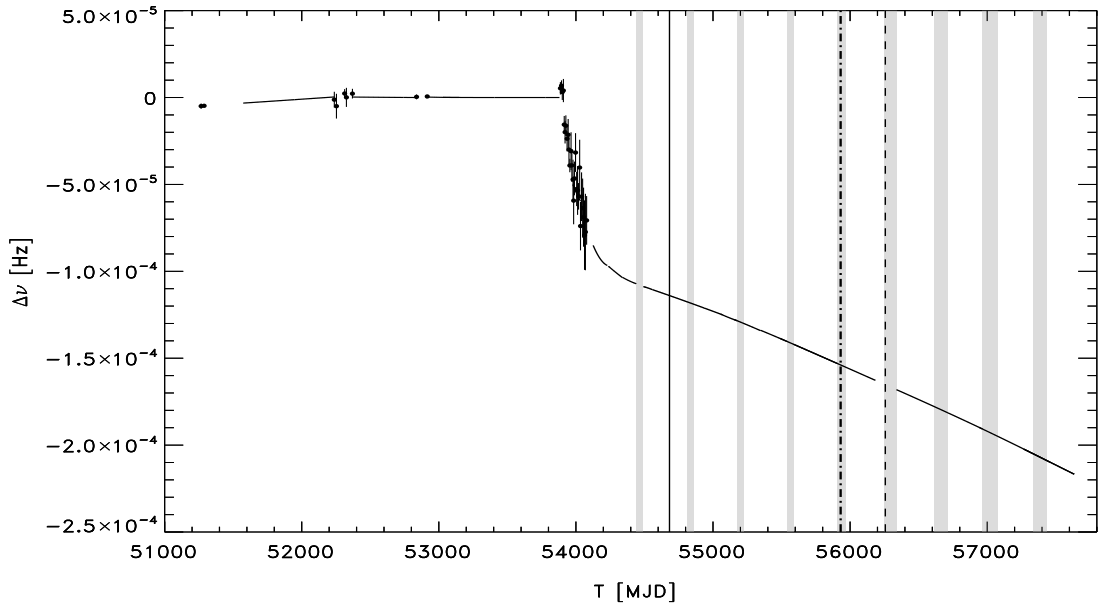}
     \caption{\label{psrj1846evol} Rotation behaviour of \psra\ from its first \rxte\ detection at 1999 April 18 up to 2016 September 4. The spin frequency 
     (solid lines represent phase coherent timing models, while data points denote incoherent measurements) is shown 
     with respect to the last phase coherent pre-outburst ephemeris \citep[MJD 53464 - 53880; see entry 5 of table 2 of][]{kuiper2009}. The solid, dashed-dotted 
     and dashed vertical lines indicate the times of the start of nominal science operations of the \fermi\ LAT instrument at MJD 54682.655 (2008 August 4) , 
     the decommisioning of \rxte\ at MJD 55931 (2012 January 5) and the start of default TTE mode operation of the \fermi\ GBM at MJD 56257 (2012 November 26). 
     The shaded bands coincide with data gaps due to observational constraints for either \rxte\ or \swift.} 
  \end{center}
\end{figure*}

\section{Timing analysis}
A timing analysis starts, irrespective the high-instrument involved, with the conversion of (selected) event arrival times registered at the satellite to
arrival times at the Solar system barycentre. This process uses the instantaneous spacecraft ephemeris (position and velocity) information, the JPL Solar 
system ephemeris information (DE200) and an accurate source position \citep[see][]{helfand2003} to convert the recorded satellite times\footnote{The on-board-registered 
event-time stamps of the (selected) events are corrected for known instrumental (fixed), ground station(s) and general time delays in the on-board-time versus 
Terrestrial-Time (TT) correlation and internal clock/oscillator-drifts (fine clock corrections).} from Terrestial Time-scale (TT or TDT, which 
differs from Coordinated Universal Time (UTC) by a number of leap seconds plus a fixed offset of 32.184 s) into Barycentric Dynamical Time (TDB) scale, 
a time standard for Solar system ephemerides. 

\subsection{Phase-coherent timing models}
Accurate timing models (ephemerides) are required to describe every revolution of the spinning neutron star accurately, otherwise a potential pulsed signal
is washed-out. These models are often composed of a limited number of Taylor series expansion components around epoch time $t_0$ in terms of spin frequency $\nu$,
frequency derivative $\dot\nu$, second order derivative $\ddot\nu$ etc.

\subsubsection{RXTE PCA}
\label{rxte_timing}
In this work we have used post-outburst \rxte\ PCA monitoring observations of \psra, covering the period 2007 August 28 till 2011 December 11 (the 
last \rxte\ observation of \psra\ before decommisioning at 2012 January 5) i.e. MJD range 54340--55906, to construct these so-called phase coherent ephemerides.
The method, which is extensively described in section 4.1 of \citet{kuiper2009}, is based on Time-of-Arrival (ToA) determinations involving a high-statistics pulse 
profile of \psra\ as correlation template. The \rxte\ monitoring based ephemerides are listed in Table \ref{eph_table} as entries 1--6.  

\subsubsection{Swift XRT}
\label{swift_timing}
In anticipation of the decommisioning of \rxte\ in 2012 January \swift\ XRT started monitoring \psra\ on  2011 July 25, and so there is about 5 months overlap in
\rxte\ and \swift\ monitoring. In this work we used the (still ongoing) \swift\ XRT monitoring observations up to and including 2016 September 3 i.e. \swift\ observations
00032031001 -- 00032031148 covering the period MJD 55767 -- 57635. In total, 147 XRT observations in WT mode have been processed and analysed, totalling an exposure time 
of $\sim 876$ ks.

Because of the spectral hardness of the pulsed emission of \psra\ in the X-ray band long exposure times, typically lasting for 15-20 ks, are required for \swift\ XRT
to detect the pulsed signal (in WT mode) at $\ga 3\sigma$ confidence levels at energies above $\sim 2.5$ keV, selecting Grade 0 events from a $30\arcsec$ aperture around 
the position of the \psra\ X-ray counterpart. These exposures can not be accommodated in a single observation given other observational constraints, and therefore we (initially) 
combined observations less than 5--10 d apart.

Initially, we attempted to augment the 2011 \rxte\ PCA ToA's with \swift\ XRT based ToA's. Inspite the poor quality of the latter ToA's we were able to add six new \swift\
ToA's, bridging the 2011--2012 data gap due to \swift\ and \rxte\ observational constraints, and to construct a combined \rxte\ / \swift\ phase-coherent timing model up to
and including 2012 September 15, covering the range MJD 55588--56185 (see entry 7 in Table \ref{eph_table}).

In the period 2012 October 6 -- November 15 three additional XRT observations were taken, lasting each $\sim 9$-$10$ ks, for which $3$-$4\sigma$ pulsed-signal significances were
found; however, phase connection with previous ToA's failed. The loss of coherence is also reported in \citet{archibald2015} and is possibly due to a glitch.

\begin{table*}
\caption{Characteristics of \integral\ observations of \psra\ for the 2008--2016 period}
\label{table:integral}
\centering
\begin{tabular}{c c c c c c c}
\\\hline\hline
Revs.    & Date begin & Date end   & MJD         & GTI$^a$ exposure & Effective$^b$ exposure & \# Scw$^c$ \\
         &            &            &             &     (Ms)     &     (Ms)      &        \\
\hline
\multicolumn{7}{c}{\textit{Post-outburst observations: 2008--2011}}\\
\\
0655-0741  & 23-01-2008  & 07-11-2008  & 54488-54777  & 1.7926       & 0.7265        & 759\\
0782-0865  & 10-03-2009  & 13-11-2009  & 54900-55148  & 1.6514       & 1.0019        & 849\\
0899-0988  & 22-02-2010  & 17-11-2010  & 55249-55517  & 1.3718       & 0.8727        & 648\\
1025-1106  & 06-03-2011  & 03-11-2011  & 55626-55868  & 0.7924       & 0.2831        & 334\\
\hline
0655-1106  & 23-01-2008  & 03-11-2011  & 54488-55868  & 5.6082       & 2.8842        &2590\\
\\
\multicolumn{7}{c}{\textit{Post-outburst observations: 2012--2015}}\\
\\
1145-1235  & 29-02-2012  & 24-11-2012  & 55986-56255  & 0.8281       & 0.3499        & 432\\
1265-1351  & 22-02-2013  & 08-11-2013  & 56345-56604  & 0.9050       & 0.4327        & 406\\
1386-1480  & 20-02-2014  & 29-11-2014  & 56708-56990  & 1.0170       & 0.5802        & 382\\
1508-1611  & 16-02-2015  & 18-11-2015  & 57069-57344  & 0.9055       & 0.5092        & 420\\
\hline
1145-1611  & 29-02-2012  & 18-11-2015  & 55986-57344  & 3.6556       & 1.8720        & 1640\\
\\
\multicolumn{7}{c}{\textit{Post-outburst observations: 2008--2015}}\\
\hline
0655-1611  & 23-01-2008  & 18-11-2015  & 54488-57344 & 9.2638       & 4.7562         & 4230\\
\hline
\multicolumn{7}{l}{{\it Notes.} $^a$ Total Good-Time-Interval exposure of the used observations}\\
\multicolumn{7}{l}{$^b$ Effective exposure on \psra\ corrected for off-axis sensitivity reduction}\\
\multicolumn{7}{l}{$^c$ Number of used Science Windows, see Sect. \ref{instr_integral}}\\
\end{tabular}
\end{table*}

For the 2013-and-beyond \swift\ XRT observations, we developed and employed a new strategy for the construction of phase coherent timing models, given the poor quality of
the ToA's and so questioning the reliability of the ephemerides. In the new method the $Z_2^2$-test statistics \citep{buccheri1983}\footnote{The \rxte\ PCA pulse-phase distribution for the $\sim 2$ - $30$ keV band shows that a fundamental plus of one harmonic provides the best description for the pulse profile.}, which is a function of the timing 
parameters $(\nu,\dot\nu,\ddot\nu)$ for given epoch $t_0$ through the pulse phase $\Phi$ of each selected event, is maximized using a three dimensional {\tt Simplex} 
optimization scheme \citep[see e.g. chapter 10.4 of][]{pressteukolsky}.
Using proper start values and scales for the model parameters e.g. those from a previous coherent solution, the scheme converges to the global maximum of $Z_2^2$.
We applied this method to the \swift\ XRT observations taken between MJD 56338--56966 (2013 February 15 -- 2014 November 5; ending just before the 2014--2015 data gap), for which 
an equivalent ToA-based ephemeris exists \citep[see][Table 1; second solution]{archibald2015}. The solution is shown in Table \ref{eph_table} as entry 8, and the parameters are
fully consistent with those shown in \citet{archibald2015}. In this manner we extended the phase-coherent ephemerides set of \psra\ with two new (partially overlapping) entries
by adding \swift\ XRT observations from 2015 February 25 to 2016 September 4 (Table \ref{eph_table} entries 9 and 10). It was also possible to phase connect across the 2014--2015 and 
2015--2016 data gaps. 

In Table \ref{eph_table} all phase-coherent post-outburst ephemerides of \psra\ are summarized, valid from MJD 54340 up to 57635 (2007 August 28 -- 2016 September 4). 
The RMS column indicates the mean deviation of the model and data (in phase unit) for the ($\chi^2$-based) ToA methods, whereas the phase-zero, $\Phi_0$, column specifies 
the offset value to be applied to the phase calculation (see equation \ref{eq:folding}) of every selected event to obtain consistent alignment with the master template.

A graphical representation of the evolution of the rotation frequency of \psra\ since its discovery with respect to the last pre-outburst ephemeris is shown in 
Fig. \ref{psrj1846evol}. \rxte\ and \swift\ data gaps are indicated by shaded vertical bands, whereas some important dates for this analysis are shown as solid (start nominal \fermi\ 
science operations), dashed-dotted (decommisioning of \rxte) and dashed (start \fermi\ GBM TTE data taking mode) lines. The loss of phase coherence between MJD 56185 and 56338 during
the \swift\ XRT monitoring period is clearly visible in this plot (note that \fermi\ GBM starts in this period its TTE data taking mode). It is also clear that the post-outburst 
spin behaviour, while smoothly evolving, is not recovering to its pre-outburst characteristics, indicating permanent changes in the rotation behaviour after the 2006 June outburst/glitch.

\subsubsection{Verification of the phase-coherent timing models using INTEGRAL ISGRI and Fermi GBM}
\label{modver}
We cross checked the validity of the \rxte\ PCA and \swift\ XRT based ephemerides listed in Table \ref{eph_table} by making pulse-phase 
distributions through pulse-phase folding the selected PCA or XRT events i.e. converting the event barycentered arrival time $t$ to 
pulse-phase $\Phi$ according to 
\begin{equation}
\Phi(t)=\nu\cdot (t-t_0) + \frac{1}{2}\dot\nu\cdot (t-t_0)^2+\frac{1}{6}\ddot\nu\cdot (t-t_0)^3 - \Phi_0 \label{eq:folding}
\end{equation} 
and subsequently sorting the pulse-phases in histograms (see e.g. the \swift\ XRT pulse-phase histogram in the top panel of 
Fig. \ref{psrj1846collage} for energies between 2.5 and 10 keV combining all available data). 

However, in this procedure we use exactly the same window function, reflecting the observation time-line of the source
for the involved instrument, as used in the generation of the models and so hidden inconsistencies could still be present.
Therefore, we searched for independent data taken by high-energy instruments on different spacecrafts using a completely 
different observation time-line. 

For this purpose we have used \integral\ ISGRI data collected on \psra\ during the 2008--2015 period
(covering \integral\ revolutions 655 -- 1611; 2008 February 23 -- 2015 November 18)\footnote{During 2016 and up to 2016 December 6, no \integral\ observations have been performed
with \psra\ within $14\fdg5$ from the pointing axis.}, and the (continuous) all-sky GBM NaI detectors aboard \fermi\ for the period 
2012 November 26 -- 2016 September 4. 
The observation log details for the \integral\ observations used in this work are shown in Table \ref{table:integral}.

\begin{figure}
  \begin{center}
     \includegraphics[width=8cm,height=12cm,angle=0,bb=120 195 420 605]{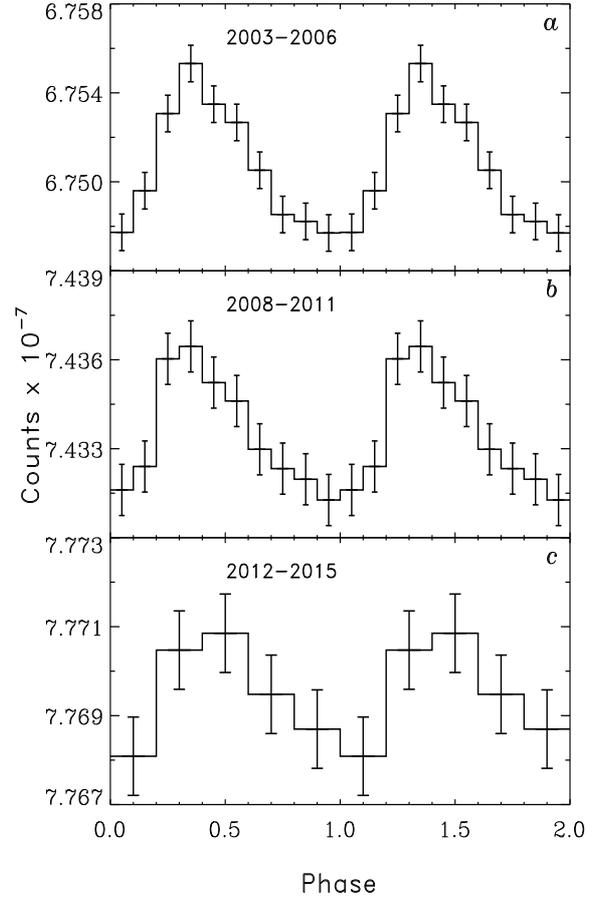}
     \caption{\label{psrj1846isgriprof} Pulse profiles of \psra\ for the 20--150 keV band as measured by \integral\ ISGRI during the pre-outburst epoch \citep[2003-2006; 
     upper panel; see fig. 3 of][]{kuiper2009} and two different post-outburst epochs (middle panel, 2008-2011; lower panel, 2012-2015) analysed in this work.}
  \end{center}
\end{figure}

The folding results of post-outburst ISGRI data (20--150 keV) for the 2008--2011 (using \rxte\ based entries 2--6 of Table \ref{eph_table}) and 2012--2015 (using \swift\ based 
entries 7--10 of Table \ref{eph_table}) epochs are shown in the middle and lower panels of Fig. \ref{psrj1846isgriprof}, respectively. The top panel of Fig. \ref{psrj1846isgriprof} 
is adapted from Fig. 3 of \citet{kuiper2009} and refers to the pre-outburst ISGRI lightcurve. It is clear from this graph that the basic (aligned) structure of the pulse-profile
of \psra\ is clearly reconstructed in the folding process for the 2008-2011 post-outburst epoch, proving the validity of ephemeris entries 2 -- 6 of Table \ref{eph_table}.
However, the measured pulsed-signal significance $Z_1^2$ of $5.9\sigma$ is lower than expected (c.f. the pre-outburst pulse-profile shown in the top panel of 
Fig. \ref{psrj1846isgriprof} has a $9.6\sigma$ significance for an effective exposure of 2.9979 Ms comparable to the 2008-2011 post-outburst value of 2.8842 Ms). 
This is partially due to an increased background level for the 2008--2011 set of mosaicked/combined observation pointings, which differs from the pre-outburst one.

We investigated the $Z_1^2$-test statistics of the 2008-2011 post-outburst and 2003--2006 pre-outburst epochs further by simulating pulse-phase distributions 
for the 20--150 keV band adopting various input strengths for the pulsed signal superposed on a (huge) dominating flat background. It turns out that, assuming a constant pulsed 
signal strength during the 2003--2006 and 2008--2011 epochs, a `genuine' (parent) pulsed flux of $\sim 85\%$ of the {\it measured} 2003--2006 epoch flux (i.e. just $1.5\sigma$ lower than measured) can statistically just explain both the $9.6\sigma$ and $5.9\sigma$ $Z_1^2$-measured values for the 2003--2006 (in the positive tail of the $Z_1^2$ distribution) and 2008--2011 (in the negative tail of the $Z_1^2$ distribution) epochs, respectively. Therefore, time variability (flux reduction during the first post-outburst epoch) of the 20--150 keV pulsed flux 
can statistically not be claimed.

For the second epoch, 2012--2015, employing only \swift\ XRT based ephemerides, the shape consistency and alignment of the 20-150 keV ISGRI profile (see bottom  panel of 
Fig. \ref{psrj1846isgriprof}) demonstrates the likely validity of the used ephemerides (entries 7--9 of Table \ref{eph_table}). However, the achieved pulsed-signal significance 
of merely $\sim 2.4\sigma$ is again (much) lower than expected, even when taking into account the considerably reduced effective exposure time on \psra\ of $1.872$ Ms for this epoch compared 
to the earlier epochs. In this case simulations demonstrate that the 20--150 keV pulsed flux should be $\la 60\%$ of the {\it measured} pre-outburst flux to explain the small $Z_1^2$ 
significance of $2.4\sigma$. This would indicate that the pulsed hard X-ray/soft \gr-ray emission {\it has} faded during the post-outburst period. A deeper study on this issue 
involving other high-energy instruments like \swift\ BAT, \fermi\ GBM or \nustar, is required, but this is outside the scope of this work.

\begin{figure}
  \begin{center}
     \includegraphics[width=8cm,height=12cm,angle=0,bb=120 195 420 605]{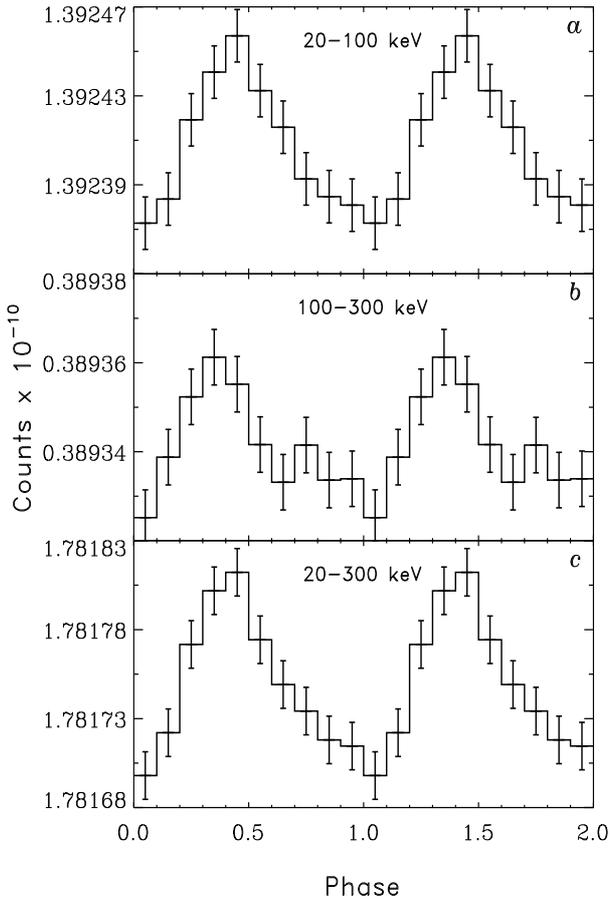}
     \caption{\label{psrj1846gbmprof} Pulse profiles of \psra\ as measured by the twelve \fermi\ GBM NaI detectors in three energy bands: 20-100 keV (top),
     100-300 keV (middle) and 20-300 keV (bottom). Data collected during mission weeks 246--430 (MJD 56337--57631; 2013 February 14 -- 2016 August 31) has been
     used in the folding process. The $Z_2^2$ significances are $6.4\sigma, 4.2\sigma$ and $7.6\sigma$ for the distributions shown in the top, middle and bottom panels, respectively.}
  \end{center}
\end{figure}

Fortunately, for the 2012--2015 epoch \fermi\ GBM TTE data have come available as of 2012 November 26, providing another way to validate the timing models. 
We have used  in the event folding procedure only TTE data from the 12 NaI-detectors. Since we are not dealing with imaging detectors, event selections
can only be made on observational conditions like imposing constraints on the following: a) source - detector pointing angle $\alpha$; b) Earth zenith - detector pointing angle $\zeta$;
c) Source - Earth zenith angle $\Psi$ and the period when the spacecraft is within/near the South Atlantic Anomaly (SAA).

In this work we applied the following maximum angles of $58\degr, 128\degr$ and $105\degr$ for selection on $\alpha,\zeta$ and $\Psi$, respectively.
The values for the first two maximum angles, valid for energies between $\sim 12$ -- $100$ keV, have been determined using ({\tt CTIME}) data on high-mass X-ray binary 
pulsar Her X-1 as `calibration' source. Above $\sim 100$ keV the source - detector pointing angle $\alpha$ widens up to $84\degr$. The maximum value for the $\Psi$ angle 
ensures that the source is never
blocked by the Earth disc which has an angular extension of about $70\degr$ as viewed from the orbit of the spacecraft. Finally, we extended the anticipated SAA duration by $\pm 300$ s 
at egress and ingress to avoid further periods of background activation.

Because the full $2\umu$s time resolution is not necessary, given the pulse period of $P \sim 328$ ms of \psra, to speed up the barycentering process we first produce lightcurves 
(counts versus TT time) in 10 ms bins for each of the 128 spectral channels of each NaI detector and convert these lightcurve TT (mid bin) times to TDB times, and these subsequently to 
pulse-phases. We have used data from the twelve Sodium Iodide detectors, which have been collected for \fermi\ mission weeks 246--430 i.e. MJD 56337--57631; 
2013 February 14 -- 2016 August 31\footnote{Because of phase coherence loss for the period MJD 56185-56338 (see Section \ref{swift_timing}) we could not fold the GBM data collected 
from the start of the TTE data taking mode at 2012 November 26 and 2013 February 15 (MJD 56257--56338; mission weeks 235--245).}.
The total GBM (NaI detectors) exposure with \psra\ in the field of view under the applied observational constraints is, averaged per detector, 19.689 Ms (32.6 weeks equivalent).

The folding process using (\swift\ XRT based) ephemerides 7--10 (see Table \ref{eph_table}) yielded the pulse-phase distributions (10 bins per cycle) shown in Fig. \ref{psrj1846gbmprof}.
The expected profile shape and alignment of \psra\ is nicely revealed in this plot, yielding for the first time even a significant detection in the 100--300 keV band of $4.2\sigma$ 
applying a $Z_2^2$ test. These results prove the validity of the \swift\ XRT based ephemerides 7--10 listed in Table \ref{eph_table}.

\begin{figure}
  \begin{center}
     \includegraphics[width=8cm,height=10cm,angle=0,bb=120 195 420 605]{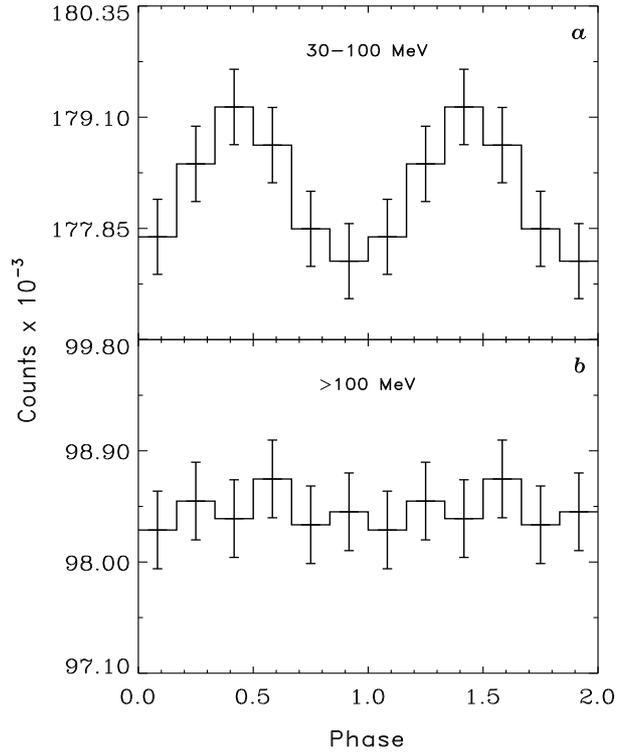}
     \caption{\label{psrj1846latprof} Pulse profiles of \psra\ as measured by \fermi\ LAT in two different energy bands, 30--100 MeV (top panel) and
     $>$ 100 MeV (bottom panel). Significant pulsed emission at $4.2\sigma$ confidence level has been detected in the 30--100 MeV band.}
  \end{center}
\end{figure}

\section{\fermi\ LAT timing and spectral results}
The validity of the newly generated timing models paved the way to proceed with a timing analysis of the \fermi\ LAT high-energy \gr-ray ($>$30 MeV) data. 
For this purpose we downloaded all Pass-8 \fermi\ LAT \citep[see e.g.][]{atwood2013a,atwood2013b,laffon2015} events registered since the start of the nominal 
\fermi\ science operations at 2008 August 4 till 2016 September 1 (MJD 54682.75--57632) from a circular Region-of-Interest (ROI) of radius $11\degr$ around \psra. 
The events were barycentered using \fermi\ tool {\tt gtbary} selecting the DE200 Solar system ephemeris file and adopting the \cxo\ X-ray position of \psra\ 
\citep{helfand2003} as best location. Next, in the event selection process using \fermi\ tool {\tt gtselect} we applied for the source analysis the following 
filters on source class and type of $evclass=128$ and $evtype=3$, respectively, as recommended for Pass 8 LAT data analysis. Also, a maximum Earth Zenith angle 
$\zeta_{Earth}^{\hbox{\rm \scriptsize max}}$ of $105\degr$ was allowed for every event arrival direction. 

\begin{figure}
  \begin{center}
     \includegraphics[width=8cm,height=6cm,angle=0,bb=65 225 535 600]{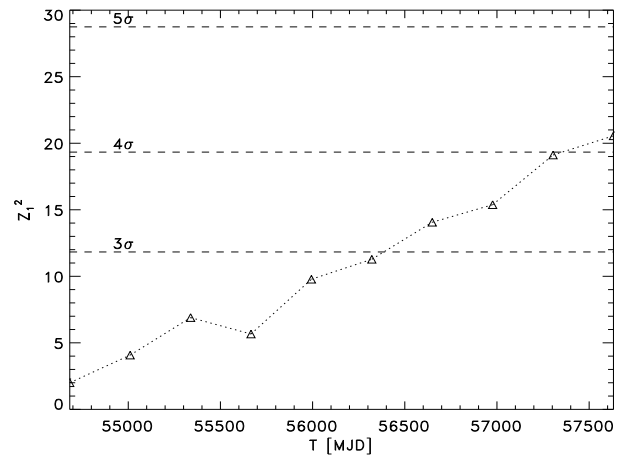}
     \caption{\label{psrj1846latznevol} The evolution of the pulsed signal strength of \psra\ in the LAT 30--100 MeV band using a stepsize of about 
     327 d. The dashed lines indicate the $3, 4$ and $5\sigma$ single trial confidence levels, from the bottom to top, respectively. The pulsed signal, $Z_1^2$, gradually reached 
     a $\sim 4.2\sigma$ significance in a linear way as expected for a constant pulsed fraction.}
  \end{center}
\end{figure}

\begin{figure}
  \begin{center}
     \includegraphics[width=8cm,height=12cm,angle=0,bb=115 145 440 650]{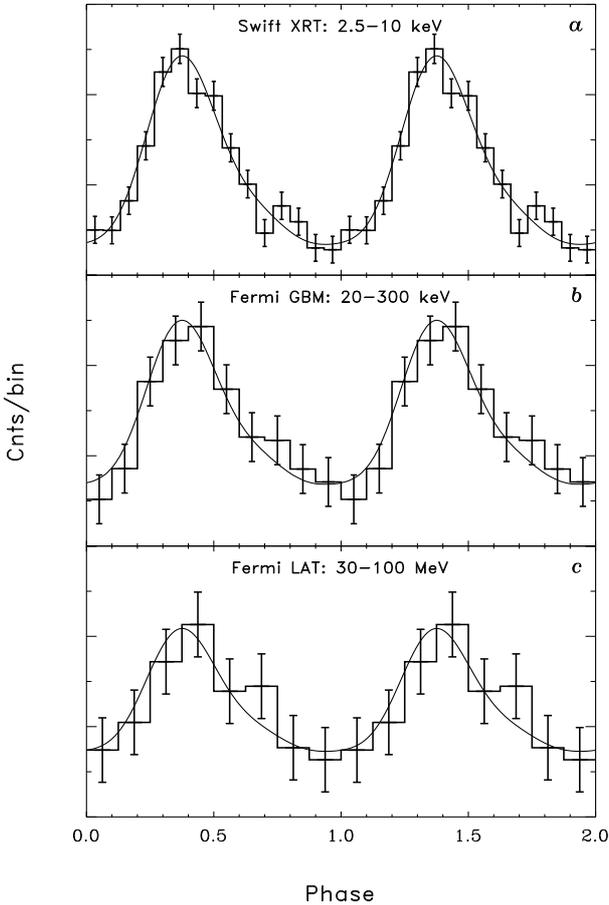}
     \caption{\label{psrj1846collage} Pulse profile collage of \psra\ covering a broad energy range from 2.5 keV to 100 MeV.
              \swift\ XRT (2.5-10 keV; top panel), \fermi\ GBM (20-300 keV; middle panel) and \fermi\ LAT (bottom panel; 30-100 MeV) profiles
              are included. Superposed in each panel is the best \integral\ ISGRI profile shape on top of a flat background. }
  \end{center}
\end{figure}
\begin{table*}
\caption{\fermi\ LAT (Pass8) pulsed excess counts, exposure and pulsed flux values of \psra\ and \msh\ for different energy bands}
\label{psrj1846pflux}
\centering
\begin{tabular}{c c c c c l}
\\\hline
$E_{-}$    & $E_{+}$ & Pulsed    & $\Gamma$    & $T_{exp}$  & \ \ \ \ \ \ \ \ \ \  Pulsed Flux\\
  (MeV)      &  (MeV)    & Counts    &             & (cm$^2$s)  & (photons cm$^{-2}$s$^{-1}$MeV$^{-1}$) \\
\hline
\multicolumn{6}{c}{\psra}\\
\vspace{-0.25cm}\\
30           & 100      & $4545 \pm 1125$ & $-3.13$ & $2.43305\times 10^{10}$ & $(3.91\pm0.97)\times 10^{-9}$  \\
100          & 1000     & $253  \pm 946$  & $-4.48$  & $1.20448\times 10^{11}$ & $(0.34\pm1.28)\times 10^{-11}$  \\
\vspace{-0.25cm}\\
\multicolumn{6}{c}{\msh}\\
\vspace{-0.25cm}\\
30           & 50      & $7116 \pm 884$ & $-2.92$ & $1.80564\times 10^{10}$ & $(2.89\pm0.36)\times 10^{-8}$  \\
50           & 70      & $4990 \pm 613$ & $-3.14$ & $4.49307\times 10^{10}$ & $(8.13\pm1.00)\times 10^{-9}$  \\
70           & 100     & $4279 \pm 536$ & $-3.34$ & $7.81044\times 10^{10}$ & $(2.68\pm0.34)\times 10^{-9}$  \\
100          & 150     & $3811 \pm 498$ & $-3.58$ & $1.27078\times 10^{11}$ & $(8.79\pm1.15)\times 10^{-10}$  \\
150          & 300     & $4138 \pm 489$ & $-3.99$ & $1.92914\times 10^{11}$ & $(2.09\pm0.25)\times 10^{-10}$  \\
300          & 500     & \, $828  \pm 264$ & $-4.50$ & $2.66704\times 10^{11}$ & $(2.27\pm0.72)\times 10^{-11}$  \\
500          &1000     & \, $239  \pm 179$ & $-5.13$ & $3.18729\times 10^{11}$ & $(0.22\pm0.17)\times 10^{-11}$  \\
1000         &10000    & $146  \pm  89$ & $-7.51$ & $3.75161\times 10^{11}$ & $(6.33\pm3.86)\times 10^{-14}$  \\
\hline
\end{tabular}
\end{table*}

\subsection{LAT timing}
In the LAT timing analysis we further selected only events from within an energy-dependent acceptation cone $\Theta_{68\%}(E)$ around \psra\ according to the average of the Pass 8 
{\tt FRONT+BACK} event type point spread functions (PSF), containing 68\% of source counts from a point-source\footnote{For more details, see http://fermi.gsfc.nasa.gov/ssc/data/analysis/docu- mentation/Cicerone/Cicerone\_LAT\_IRFs/IRF\_PSF.html}. To further suppress in our sample events possibly coming from the Earth disc,
which is mainly effective for events with energies below $100$ MeV because of the broad PSF (larger than $10\degr$ for $E_{} \la 50$ MeV),
we applied an energy dependent Earth zenith selection according to $\zeta_{Earth}^{\hbox{\rm \scriptsize max}}(E)=\zeta_{Earth}^{\hbox{\rm \scriptsize max}} - N_{\sigma}\cdot \Theta_{68\%}(E)$ with $N_{\sigma}=2$.

Next, we sorted the events in two broad energy bands, 30--100 MeV and $>$100 MeV, and folded the barycentric arrival times of the selected events on the timing models shown in
Table \ref{eph_table}. Events falling within MJD 56185-56338, where phase coherence was lost, were excluded in the folding process. 

The resulting pulse-phase distributions for the 30--100 MeV and $>$100 MeV bands are shown in Fig. \ref{psrj1846latprof}. The 30--100 MeV pulse-phase distribution, shown in the 
top panel of Fig. \ref{psrj1846latprof}, deviates from uniformity, applying a $Z_1^2$-test, at a $4.2\sigma$ confidence level, representing thus the {\it first} detection 
(single trial) of pulsed emission from \psra\ at high-energy \gr-rays. Above 100 MeV (bottom panel of Fig. \ref{psrj1846latprof}), the distribution is consistent with being uniform. 
This behaviour is in line with expectations if the spectral characteristics of \psra\ are similar to those of the `canonical' soft \gr-ray pulsar \msh\ which yielded a very soft spectrum for energies
above 30 MeV \citep{kuiper2015}. It does not support a flat spectral shape (power-law index $\sim -2$ at high-energy gamma rays till a break at GeV energies) as measured for \psrb.

We investigated the evolution of the pulsed signal strength in the 30--100 MeV band as a function of integration time (see Fig. \ref{psrj1846latznevol}). 
We see a gradual linearly build-up of the pulsed signal $Z_1^2(t)$. This is expected for a genuine pulsed signal with a (nearly) constant pulsed fraction $\hat{p}$, because the following 
relation holds \citep{dejager1987}: $Z_1^2(t)=(N_{tot}(t)-1)\cdot \hat{p}^2 + 2 \label{eq:z1evol}$, with $\hat{p}=N_{pul}(t)/N_{tot}(t)$ the pulsed fraction uncorrected for 
background, $N_{tot}(t)$ the total number of selected events within the acceptance cone (= total number of events from \psra\ plus events from the dominating, mainly celestial, background) as a 
function of time and finally $N_{pul}(t)$ the number of pulsed counts from \psra\ within the acceptance cone as a function of time. 

From the \swift\ XRT (soft X-rays), \fermi\ GBM (hard X-rays/soft \gr-rays) and \fermi\ LAT (\gr-rays; 30--100 MeV) data we compiled a pulse-profile collage showing the shape of \psra\ as a function of energy (see Fig. \ref{psrj1846collage}). In this figure, the best-fitting \integral\ ISGRI profile \citep[20-150 keV; see e.g. panel a of Fig. \ref{psrj1846isgriprof} and figure 3 of ][]{kuiper2009} is superposed. There is no evidence for morphology changes of the profile as a function of energy across the energy band $\sim 2.5$ keV - $100$ MeV.

\subsection{LAT \psra\ pulsed-flux determination}
\label{subpuls1846}
Because the pulsed signal from \psra\ is detected only in a narrow bandpass below 100 MeV a detailed characterization of the high-energy \gr-ray spectrum is impossible
\footnote{The lack of a (proper) full description of the high-energy \gr-ray spectrum prevents also the use of the photon weighting method \citep{kerr2011} to improve the 
sensitivity to the weak pulsation as detected in the 30--100 MeV band.}. We can, however, estimate the 30--100 MeV pulsed photon flux of \psra\ from the measured 30--100 MeV (pulsed) excess counts $N_p$ (i.e. pulsed counts above flat background; see Fig. \ref{psrj1846collage}c) and the LAT exposure $T_{exp}$ for the 30--100 MeV band assuming a certain photon spectral index across this band. 

For the pulsed excess counts we derived a value of $4545\pm 1125$ fitting a pre-defined pulse shape, the ISGRI 20-150 keV shape \citep[cf. Fig. 3 of][]{kuiper2009}, to the measured 30--100 MeV pulse-phase histogram (see Fig. \ref{psrj1846collage}c). These counts have been accumulated across the MJD 54682.75--57632 period, excluding time interval MJD 56185-56338 for which phase coherence was lost.

The second quantity, $T_{exp}$, has been determined using \fermi\ analysis tool {\tt gtexposure}, setting {\tt P8R2\_SOURCE\_V6} as reference to instrument (Pass-8) response functions and adopting
for the weighting of the exposure as a function of energy across the 30--100 MeV band a photon spectral index of $-3.13$. This index is consistent with the value earlier derived for \msh\ and confirmed for \msh\ and \psra\  below in the fits to 
their broad-band spectra (see Section \ref{broad-band}).
The {\tt gtexposure} tool also requires a `counts' lightcurve which has been prepared
by {\tt gtbin} adopting a binsize of $\frac{1}{4}$ day.  We ended up with an exposure $T_{exp}$ of $2.43305\times 10^{10}$ cm$^2$s for the 30--100 MeV band across the MJD 54682.75--57632 period (excl. MJD 56185-56338).

Now, all ingredients are available to calculate the pulsed flux, $F_p$, from $F_p= ({N_p}/{f_{1\sigma}}) \cdot ({1}/{T_{exp}}) \cdot ({1}/{(E_{+}-E_{-})})$, in which 
${1}/{f_{1\sigma}}$ represents the correction factor for the (missing) flux outside the Pass-8 $1\sigma$ source-acceptance cone ($f_{1\sigma}=0.6827$), and $E_{-}$ and $E_{+}$ represent the lower- and upper bounds of the energy band, respectively. The resulting pulsed flux values for \psra\ are listed in Table \ref{psrj1846pflux} along with those derived for \msh\ (see Section \ref{subpuls1509}).

\begin{table*}
\caption{Phase-coherent X-ray (PCA)/radio ephemerides for \msh\ covering the time period MJD 54626-57372 (2008 June 9 -- 2015 December 16). Solar system planetary ephemeris DE200 adopted.}
\label{eph1509_table}
\begin{center}
\begin{tabular}{lccclllcc}
\hline
Entry &  Start &  End  &   t$_0$, Epoch   & \multicolumn{1}{c}{$\nu$}   & \multicolumn{1}{c}{$\dot\nu$}               & \multicolumn{1}{c}{$\ddot\nu$}                  & $\Phi_{0}$  & Validity range\\
 \#   &  [MJD] & [MJD] &     [MJD,TDB]    & \multicolumn{1}{c}{[Hz]}    & \multicolumn{1}{c}{$\times 10^{-11}$ Hz s$^{-1}$}  & \multicolumn{1}{c}{$\times 10^{-21}$ Hz s$^{-2}$}  &             &  (days)   \\
\hline\hline
1           & 54626 & 55075 & 54626.0     & 6.601\,176\,760\,2(3)  & -6.664\,902(3)          &   1.969(2)                  & 0.4573 &  450\\
2           & 55075 & 55465 & 55075.0     & 6.598\,592\,680\,2(3)  & -6.657\,350(3)          &   1.884(2)                  & 0.0151 &  391\\
3           & 55465 & 55927 & 55465.0     & 6.596\,350\,488\,4(4)  & -6.650\,918(4)          &   1.885(2)                  & 0.4970 &  463\\
4           & 55939 & 56376 & 56157.0     & 6.592\,377\,363(3)     & -6.639\,61(2)           &   1.88(7)                   & 0.3758 &  438\\
5           & 56310 & 56708 & 56509.0     & 6.590\,358\,953(4)     & -6.633\,84(3)           &   1.88(10)                  & 0.3375 &  399\\
6           & 56670 & 57081 & 56876.0     & 6.588\,256\,398(3)     & -6.627\,83(3)           &   1.94(11)                  & 0.1454 &  412\\
7           & 57013 & 57372 & 57192.0     & 6.586\,447\,550(4)     & -6.622\,67(4)           &   1.89(16)                  & 0.7781 &  360\\
\hline
\end{tabular}
\end{center}
\end{table*}

\subsubsection{LAT \msh\ pulsed-fluxes revisited}
\label{subpuls1509}
\begin{figure}
  \begin{center}
     \includegraphics[width=7cm,height=14cm,angle=0,bb=134 85 440 695]{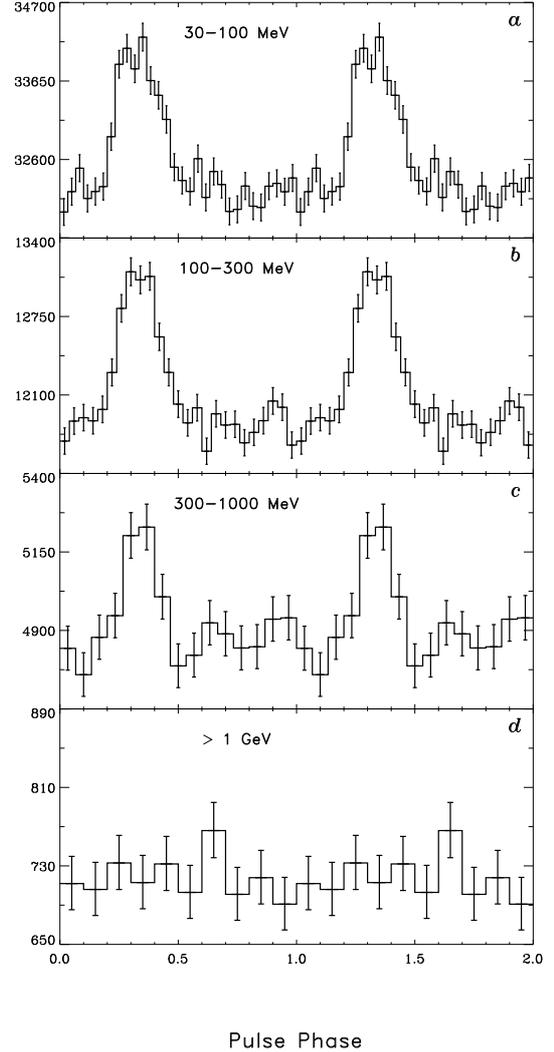}
     \caption{\label{psrb1509latprof} \fermi\ LAT pulse profiles of \msh\ in four different energy bands using Pass8 data
     collected during 2008 August 4 and 2015 December 16 (MJD 54682.85-57372). Even for the 300--1000 MeV band significant pulsed emission 
     ($5.5\sigma$) has been detected.
     }
  \end{center}
\end{figure}

The availability of \fermi\ LAT Pass-8 data, and the greatly increased exposure time of about 7.4 yr on the soft \gr-ray pulsar \msh\ triggered us to revisit the high-energy pulsed gamma-ray emission of \msh. 
This offers a verification of the earlier published results for \msh\ and allows a comparison with the derived characteristics of \psra\ using for both  Pass-8 data and response parameters. 
The latest \gr-ray spectral results for \msh\ were published by \citet{kuiper2015} using Pass-7 LAT data and a data-collecting period of 3.4 yr. They obtained a $10.2\sigma$ pulsed signal for photons 
with energies between 30 and 1000 MeV. Now, using Pass-8 data and a 7.4 yr data-collection period yields a $28.7\sigma$ pulsed signal for the 30--1000 MeV band applying appropriate and up-to-date 
X-ray/radio based ephemerides (see Table \ref{eph1509_table}) in the timing analysis. This dramatic improvement in sensitivity makes a detailed study in differential energy bands possible. Fig. \ref{psrb1509latprof} shows the (Pass-8) LAT pulse profiles for four different energy bands: 30--100 MeV, 100--300 MeV, 300--1000 MeV and $>1$ GeV, yielding pulsed signal significances ($Z_3^2$-test) of $19.5\sigma, 18.8\sigma, 5.5\sigma$ and $< 1\sigma$, respectively. Pulsed emission has been detected up to $\sim 500$ MeV!

From the pulse-phase distributions - in even smaller energy bands - we derived pulsed excess counts by subtracting the `unpulsed' level, estimated in phase interval 0.7--1.2, from the counts collected in the `pulsed' phase interval covering 0.2--0.7. The derived excess counts were converted to pulsed fluxes analogous to the method employed for \psra, as outlined in Section \ref{subpuls1846}. The resulting pulsed fluxes are given in Table \ref{psrj1846pflux}.

\subsubsection{Broad-band spectra of the pulsed emission of \psra\ and \msh}
\label{broad-band}
\begin{figure}
  \begin{center}
     \includegraphics[width=8.5cm,height=8.5cm,angle=0,bb=55 157 560 655]{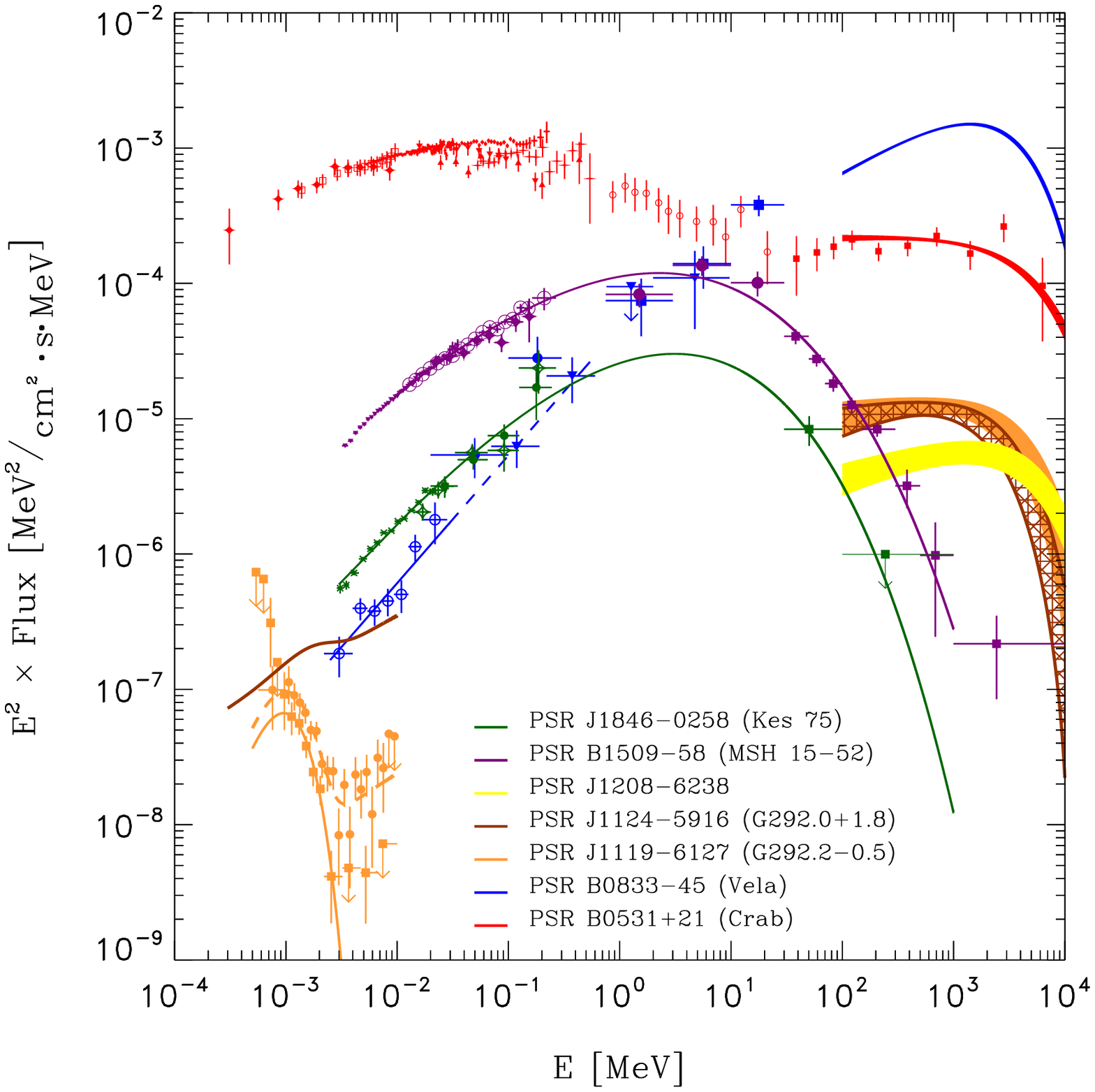}
     \caption{\label{psrj1846tpspc} The high-energy pulsed emission spectra of \psra\ (green symbols/line) and \msh\ (purple symbols/line) from $\sim 2.5$ keV up to $\sim 1000$ MeV.
     For comparison also the pulsed spectra of the Crab (red) and Vela (blue; the strongest high-energy \gr-ray source) pulsars are shown. The \gr-ray spectra of three other high B-field 
     pulsars are superposed: \psrb\ (orange), \psrc\ (dark-orange/red) and \psrn\ (yellow). The total (pulsed plus unpulsed) X-ray spectra based on \xmm\ data of \psrb\ (data points/dashed line)
     and \psrc\ are shown as well. The pulsed (thermal) component of \psrb\ is superposed as a solid orange line.}
  \end{center}
\end{figure}

The newly derived \fermi\ LAT (Pass-8) pulsed-flux values for \psra\ and \msh\ (see Table \ref{psrj1846pflux}) are plotted in Fig. \ref{psrj1846tpspc} as green- and purple data points (filled squares), respectively, along with pulsed flux measurements derived at lower energies \citep[see][for the latter values]{kuiper2015}. We fitted the broad-band pulsed emission spectra of both \psra\ and \msh\ with a model  
\citep[see Section 5.7 of][]{kuiper2015} with the following $E_{\gamma}$ dependence: 
\begin{equation}
F_{\gamma}= k \cdot (E_{\gamma}/E_0)^{\Gamma}\cdot \exp(-(E_{\gamma}/E_c)^{\beta})\label{eq:model}
\end{equation}

The normalization energy $E_0$, which minimizes the correlation between the four fit parameters, was $0.0243069$ MeV for the \psra\ spectral data set. The best-fitting values 
were $k=(2.05\pm0.03)\cdot 10^{-2}$ ph cm$^{-2}$s$^{-1}$MeV$^{-1}$; $\Gamma=-0.932\pm 0.015$; $E_c=0.0087\pm 0.0006$ MeV and $\beta=0.245\pm 0.007$. For these model parameters (and uncertainties in these) 
the maximum luminosity of the pulsed emission of \psra\ is reached at
$E_{\gamma}^{\hbox{\scriptsize{\rm max}}}=E_c \cdot (\frac{\Gamma+2}{\beta})^{\frac{1}{\beta}} = 3.5 \pm 1.1 $ MeV.

The best-fitting model parameters for \msh\ were $k=(4.81\pm0.03)\cdot 10^{-2}$ ph cm$^{-2}$s$^{-1}$MeV$^{-1}$; $\Gamma=-1.067\pm 0.003$; $E_c=0.00234\pm 0.00004$ MeV and $\beta=0.2144\pm 0.0008$, whereas $E_0$ was $0.105745$ MeV.
This results in a $E_{\gamma}^{\hbox{\scriptsize{\rm max}}}$ value of $2.23 \pm 0.11$ MeV, consistent with the value of $\sim 2.5$ MeV given in \citet{kuiper2015} and $2.6\pm0.8$ MeV by \citet{chen2016}, 
who included an accurate spectrum up to $\sim 79$ keV analysing \nustar\ data. The best-fitting models are also superposed in Fig. \ref{psrj1846tpspc} as green (\psra) and purple (\msh) solid lines. Their shapes are remarkably similar. It is clear that both pulsars reach their maximum luminosities in the MeV band. 

Comparing fig. 7 of \citet{kuiper2015} with Fig. \ref{psrj1846tpspc} of this work it is evident that the statistical quality of the LAT pulsed flux measurements of \msh\ drastically improved, going from Pass 7 to Pass 8, and more than doubling the exposure time.

\section{Summary}

For our successful attempt to detect and characterize the pulsed signal of the radio-quiet magnetar-like pulsar \psra\  in the high-energy
\gr -ray data of \fermi\ LAT, we first had to construct an ephemeris using its pulsed X-ray emission. This complex initial step was crucial
for obtaining our results.

(1) We succeeded in producing phase-coherent timing models exploiting \rxte\ PCA and \swift\ XRT monitoring data for the post-outburst period
from 2007 August 28 (MJD 54340) to 2016 September 4 (MJD 57635) (see Table \ref{eph_table}). The smoothly evolving post-outburst spin behaviour is not
recovering to its pre-outburst characteristics, indicating permanent changes in the rotation behaviour after the 2006 outburst/glitch (Fig. \ref{psrj1846evol}).

(2) Independent verification of the ephemerides was obtained using \integral\ ISGRI and \fermi\ GBM data by making pulse-phase distributions through phase folding selected ISGRI (20--150 keV)
and/or GBM (20--300 keV) events, accumulating over long time intervals (2008 February 23 -- 2016 September 4, and  2013 February 14 -- 2016 August 31, respectively). The constructed broad pulse
profiles were  identical in shape as measured post outburst by \rxte\ PCA and \swift\ XRT and by \integral\ ISGRI during the pre-outburst epoch (2003--2006), 
confirming the correctness of the ephemerides (Figs \ref{psrj1846isgriprof} and \ref{psrj1846gbmprof}). 

(3) Interestingly, we found in the multi-year \integral\ ISGRI data an indication for fading of the pulsed hard X-ray/soft \gr-ray emission during the post-outburst epoch (see Sect. \ref{modver}).

(4) Taking advantage of the increased sensitivity of \fermi\ LAT by using Pass-8 data, and phase folding barycentric arrival times of selected events
with the ephemerides produced in this work, we obtained a $4.2\sigma$ detection
of a pulsed signal for energies 30--100 MeV, with a pulse consistent in shape and aligned in phase with the profiles measured at lower energies, e.g. \swift\ XRT (2.5--10 keV), \integral\ ISGRI (20--150 keV) and
\fermi\ GBM (20--300 keV) (Fig. \ref{psrj1846collage}). 
The flux (30--100 MeV) of the broad pulse is $(3.91\pm0.97)\times 10^{-9}$ photons cm$^{-2}$s$^{-1}$MeV$^{-1}$.

(5) For energies above 100 MeV the \gr -ray pulse could not be detected, indicative for a very soft spectral shape at high-energy \gr-rays.

(6) We rederived  the timing and spectral characteristics of \msh\  at high-energy \gr -rays, exploiting the increased sensitivity of Pass-8 data and the greatly increased \fermi\ LAT exposure time 
(collected over about 7.4 yr), compared to our earlier report on this pulsar in \citet{kuiper2015}. The shape of the published broad pulse profile and high-energy spectrum have been confirmed 
with greatly improved statistics (Fig. \ref{psrb1509latprof} and Table \ref{psrj1846pflux}, respectively).

(7) The broad-band pulsed emission spectra (from 2 keV up to \fermi\ energies) of \psra\ and \msh\ can both be accurately described with similarly curved shapes with a photon energy, $E_{\gamma}$, 
dependence as given in equation (\ref{eq:model}) (Section \ref{broad-band} and Fig. \ref{psrj1846tpspc}). For the best-fitting parameters, the maximum luminosity of 
the pulsed emission of \psra\ is reached at $3.5 \pm 1.1$ MeV and for \msh\ at $2.23 \pm 0.11$ MeV.

\begin{table*}
\renewcommand{\tabcolsep}{1.8mm}
\begin{center}
\caption{All high-magnetic-field rotation-powered pulsars with characteristic age $\la$ 3 kyr detected by \fermi\ LAT}. 
 \label{fermihighbyoung}
\begin{tabular}{lcccccccl}
\hline\noalign{\smallskip}
name                          & $P$   & $\dot{P}$     & Age    &  $B_{\hbox{\scriptsize s}}$ &  $L_{\hbox{\scriptsize sd}}$     & Pulse  shape            & $E(L_{max})$ & Comment \\
                              &  (ms) &  ($10^{-12}$) &(kyr)   &      $(10^{13} $G)          &  ($10^{37}$erg s$^{-1}$)         & (\gr-rays)              & (MeV)        &         \\
\hline\noalign{\smallskip}
\noalign{\smallskip}
\psrb\ (G292.2-0.5)           & 407   & 4.02          & 1.6    & 4.1                         &  0.23                            & two pulses$^{[1]}$ & $600$        &  r,X,$\gamma$\,/magnetar-like outburst\\
\psrc\ (G292.0+1.8)           & 135   & 0.75          & 2.9    & 1.0                         &  1.19                            & two pulses$^{[2]}$ & $520$        &  r,X,$\gamma$  \\
\psrn\                        & 440   & 3.27          & 2.7    & 3.8                         &  0.15                            & two pulses$^{[3]}$ & $1300$       &  $\gamma$  \\
\msh\  (MSH 15-52)            & 151   & 1.53          & 1.6    & 1.5                         &  1.72                            & single broad       & $2.2$        &  r,X,$\gamma$  \\                
\psra\  (Kes 75)              & 326   & 7.13          & 0.7    & 4.9                         &  0.81                            & single broad       & $3.5$        &  X,$\gamma$\,/magnetar-like outburst \\
\noalign{\smallskip}
\hline
\multicolumn{9}{l}{{\it Notes}. Column 1 gives source name(s) and associated SNR or PWN (between brackets), when applicable.}\\ 
\multicolumn{9}{l}{Columns 2 and 3 give the period $P$ and the period derivative $\dot{P}$; Column 4 gives the characteristic age ($\tau=-0.5\nu/\dot{\nu}$).}\\
\multicolumn{9}{l}{Column 5 gives the magnetic-field strength at the surface $B_{\hbox{\scriptsize s}}$; Column 6 gives the spin-down power ($L_{\hbox{\scriptsize sd}}=4\upi^2I\nu\dot{\nu}$), in erg s$^{-1}$.}\\
\multicolumn{9}{l}{Column 7 gives for the \fermi\ band above 30 MeV a description of the pulse shape; [1] \citet{parent2011}, [2] \citet{abdo2010a}, [3] \citet{clark2016}.}\\
\multicolumn{9}{l}{Column 8 gives the energy where the maximum luminosity is reached in the broad-band spectrum.}\\
\multicolumn{9}{l}{Column 9 a label r, X, $\gamma$\ indicates that pulsed emission has been detected at radio wavelengths, X-rays and \gr-rays, respectively.}\\
\end{tabular}
\end{center}
\end{table*}

\section{Discussion}
Table \ref{fermihighbyoung} lists all high-magnetic-field rotation-powered pulsars with characteristic age $\la$ 3 ky detected by \fermi\ LAT. The five \gr-ray pulsars listed, have
rather similar rotation (and derived) characteristics, but exhibit very different characteristics at \gr-ray energies. 
\psrb , \psrc\ \citep{camilo2002} and \psrn\ \citep{clark2016} show two pulses 
in their \gr-ray pulse profiles and reach their maximum luminosities in the GeV band, like most of the pulsars in the \fermi\ pulsar catalog. In this discussion we call these pulsars 'GeV pulsars'.  
On the other hand, \msh\ and \psra\ have single broad \gr-ray pulses and reach their maximum luminosities at low-energy \gr-rays. These pulsars we call 'MeV pulsars'.

Interestingly, the two magnetar-like rotation-powered pulsars \psrb\ and \psra, looking like twins in their rotational parameters, appear to be widely different at X-rays and high-energy \gr-rays. 
Fig. \ref{psrj1846tpspc} shows the  broad-band (pulsed-emission for \psra\ and \msh) spectra of all five pulsars in comparison with those of the Crab and Vela pulsars. \psrb , \psrc\ and \psrn\, with their 
weak emissions at X-rays (\psrn\ not yet detected), exhibit broad-band spectra more similar to that of Vela, than that of the Crab. 
 \psra\ is now the second pulsar, after \msh, shown to reach maximum luminosity at MeV energies by {\it measuring} its broad-band spectrum into the \fermi\ LAT band of high-energy \gr-rays.

\citet{kuiper2015} presented the soft \gr-ray pulsar catalogue containing 18 non-recycled rotation-powered pulsars from which non-thermal pulsed emission has been securely detected 
at hard X-rays/soft \gr-rays above 20 keV. The majority (11 members) exhibits broad, structured single-pulse profiles at soft \gr-rays, like \psra\ and \msh\, and 15 show hard 
power-law spectra in the hard X-ray band. \msh\ being the exception, \fermi\ LAT had not yet detected the pulsed emission from the remaining 14. Given the high sensitivity of 
the LAT above 100 MeV, it was concluded that these 14 pulsars, including \psra, also had to reach their maximum luminosities typically at MeV energies. 
Interestingly, the soft \gr-pulsars are all fast rotators and on average an order of magnitude younger and $\sim40$ times more energetic than the \fermi\ LAT sample, suggesting 
that we are dealing with a special subset of high-energy rotation-powered pulsars. This makes it particularly interesting to decisively establish their manifestation as MeV pulsars. 
For \psra\ this has been succeeded in this work.

There is still no concensus on the physics behind the deviant curved broad-band spectral shape of the MeV pulsars. This has been extensively discussed for \msh\ after the broad-band 
high-energy spectrum was revealed by  the instruments aboard the Compton Gamma-Ray Observatory (\cgro), summarized in \citet{kuiper1999}. In the latter paper, the firm detection of 
pulsed emission at MeV energies by \cgro\ COMPTEL was reported, together with low upper limits/weak detections with \cgro\ EGRET below 100 MeV. \citet{harding1997} proposed for the 
Polar Cap (PC) model, where the gamma rays are produced near the stellar surface, the quantum electrodynamic process of magnetic photon splitting, 
as an explanation for the apparent lack of detection at GeV energies.
This process becomes important when the pulsar magnetic field near the surface approaches the quantum critical value $B_{cr} = 4.41 \times 10^{13}$ G.
They concluded that photon splitting, or combined splitting and pair production, can explain the broad-band spectrum of \msh\ with the unusually low cutoff energy. Alternatively, 
\citet{zhang2000} reproduced the cutoff in the high-energy spectrum as well as the broad pulse profile of \msh\ in the context of the Outer Gap model (OG) model, where the \gr-ray 
radiation comes from the outer magnetosphere. The non-thermal photons are emitted by $e^{\pm}$ pairs produced by back-flow charged particles from the OG through synchrotron radiation 
mechanism near the stellar surface and in a finite region just above the OG through a synchrotron self-Compton mechanism.

The broad-band spectral shape of \msh\  as presented by \citet{kuiper1999} was confirmed by \citet{pilia2010} and \citet{abdo2010} using the next-generation of gamma-ray instruments 
{\it AGILE} and \fermi, respectively. \citet{abdo2010} used only 1 yr of survey data with \fermi\ LAT and detected the pulsed signal of \msh\ at the 3$\sigma$ level in the energy 
intervals 30--100 MeV and 100--300 MeV. This was obviously a small fraction of the 7.4 years of survey data used in this work. \citet{kuiper1999}, \citet{pilia2010} and \citet{abdo2010} 
extensively discuss OG, Slot Gap  and PC scenarios (see references therein) to explain the spectral and temporal characteristics of \msh\, and conclude for example, that the PC model  
including photon splitting is spectroscopically viable, but subject to the strong constraint of emission at the magnetic co-latitude of the rim, i.e. $\sim 2^{\circ} $ as proposed 
by \citet{kuiper1999}.

More recently, \citet{wang2013} proposed a new version of the OG model to successfully explain the characteristics of \msh. They explain that most pairs created in the OG are 
created around the null charge surface and the gap's electric field separates the two charges to move in opposite directions. The region towards the light cylinder is dominated by 
the outflow current, producing curvature radiation as measured from the \fermi\ LAT GeV pulsars.
The region towards the neutron star is dominated by the inflow radiation, in which magnetic pair creation converts curvature photons into pairs by the strong magnetic field. 
The hard X-rays and soft gamma rays of \msh\ result from synchrotron radiation of these pairs. They argue that the observer viewing angle measured from the rotation axis is smaller 
than (or close to) the inclination angle of the magnetic axis. For this geometry, the outward GeV emissions are missed by the observer, while the inward emissions are observed. 

In a follow-up paper \citet{wang2014} applied this scenario to four young pulsars,
including \psra\ with PSR J1617--5055, PSR J1811--1925 and PSR J1930+1852. 
These form a subset of a preliminary version of the soft \gr-ray pulsar catalog \citep{kuiper2015} and share some emission properties with \msh\ : (1) their radio emissions 
are dim or quiet; (2) the pulse profile in X-rays/soft \gr-rays is described by a single broad profile; (3) no GeV emissions have been detected; (4) the broad-band spectral shape 
suggests that they are all MeV pulsars.
\citet{wang2014} underlined that the viewing geometry is a crucial factor to discriminate between the normal GeV pulsars and the MeV pulsars.
Furthermore, the magnetic inclination angle  of the MeV pulsars is relatively small, $\alpha \le 30^{\circ} $. For all four pulsars, including \psra, and for \msh\ \citep{wang2013}, 
broad-band spectra and pulse profiles were calculated that match the observed characteristics. 

As we mentioned above, in the hard X-ray/soft \gr-ray pulsar catalogue \citep{kuiper2015}, the number of (candidate) MeV pulsars has increased to eleven, constituting a fraction of 
about 60\%. This underlines the importance of understanding the physics behind this manifestation: Are we seeing GeV and MeV pulsars due to, for example, differences in viewing directions and
geometries in OG models as proposed by \citet{wang2014}, or plays the signature of the exotic process of photon splitting in a very strong magnetic field near the stellar surface a
dominant role \citep{harding1997}? 
The latter scenario, however, can not explain the characteristics of all MeV pulsars, since some of them, including PSR J1617--5055 and PSR J1811--1925, appear not to posses sufficiently
strong magnetic fields approaching the quantum critical value. Furthermore, we show in Fig. \ref{psrj1846tpspc} that \fermi\ LAT has detected high B-field pulsars, including the 
magnetar-like \psrb, as GeV pulsars with their maximum luminosities at GeV energies. For the latter arguments we are inclined to believe more in geometrical solutions to the problem 
of understanding the scenarios that explain the manifestations of the young and energetic MeV pulsars.

In this work we probed in detail the spectral extension of the MeV pulsars \psra\ and \msh\ in the \fermi\ LAT band, going to the bottom for \psra. Other promising MeV-pulsars
to show spectral extensions in the GeV band are \axj\ and \igr, each having stronger pulsed emission at hard X-rays/soft \gr-rays than \psra\ 
\citep[see fig. 28 of][]{kuiper2015}).
Both pulsars, however, are not detected at radio wavelengths, and could only be timed at X-rays. \axj\ is currently (since 2017 March 19) being monitored by \swift\ XRT to obtain 
phase coherent timing models that will be used in turn to uncover a likely pulsed gamma-ray signal in a timing analysis of \fermi\ LAT data. We already have folded \fermi\ LAT Pass-8 
events collected during 2008 Aug. 4 and 2010 December 12, when \rxte\ PCA monitored the source \citep[see table 1 of][for the phase coherent timing models]{kuiper2015}, and this yielded, 
inspite the low exposure of only 2.2 yr, encouraging $2.1-2.4\sigma$ pulsed signal significances for the 30--1000 MeV band. 

Future high-sensitivity MeV telescopes like the proposed AMEGO or e-ASTROGAM missions are required to make significant progress in understanding the physics of MeV pulsars.


\section*{Acknowledgments}
This research has made use of data and/or software provided by the High Energy Astrophysics Science Archive Research Center (HEASARC), 
which is a service of the Astrophysics Science Division at NASA/GSFC and the High Energy Astrophysics Division of the Smithsonian Astrophysical 
Observatory. We acknowledge the use of public data from the \swift\ and \fermi\ data archives. 
In this work we analysed observations performed with \integral, an ESA project with instruments and science data centre funded by ESA member 
states (especially the PI countries: Denmark, France, Germany, Italy, Switzerland, Spain) and with the participation of Russia and the USA.
We also acknowledge the use of NASA's Astrophysics Data System (ADS) and of the SIMBAD data base, operated at CDS, Strasbourg, France.
We thank Dr. David Smith for useful discussions on \fermi\ LAT pulsar timing issues, and Dr. Matthew Kerr for providing radio pulsar ephemerides for \msh.

\bsp

\label{lastpage}

\end{document}